\documentstyle[sprocl]{article}

\input{epsf}
\def\Journal#1#2#3#4{{#1} {\bf #2}, #3 (#4)}

\def\NPB{{\em Nucl. Phys.} B}
\def\PLB{{\em Phys. Lett.}  B}

\def\PRD{{\em Phys. Rev.} D}

\def\be{\begin{equation}}
\def\ee{\end{equation}}
\def\bea{\begin{eqnarray}}
\def\eea{\end{eqnarray}}

\def\msbar{\overline{\mbox{MS}}}

\newcommand{\al}{\alpha}
\newcommand{\bt}{\beta}
\newcommand{\gm}{\gamma}

\newcommand{\et}{\eta}

\newcommand{\lm}{\lambda}
\newcommand{\rh}{\rho}

\newcommand{\ta}{\tau}

\newcommand{\ph}{\phi}
\newcommand{\vr}{\varphi}

\newcommand{\om}{\omega}

\newcommand{\Sg}{\Sigma}
\newcommand{\btb}{\bar{\bt}}
\newcommand{\lmb}{\bar{\lm}}
\newcommand{\vb}{\bar{v}}

\newcommand{\vecx}{{\bf x}}

\newcommand{\half}{\frac{1}{2}}

\newcommand{\Tr}{\mbox{Tr}\,}

\newcommand{\phd}{\ph^{\dagger}}

\begin{document}

\title{NUMERICAL STUDY OF PLASMON PROPERTIES IN THE SU(2)-HIGGS MODEL}

\author{Jan SMIT}

\address{Institute of Theoretical Physics, University of Amsterdam,\\
Valckenierstraat 65, 1018 XE Amsteram, the Netherlands}




\maketitle\abstracts{
We discuss an explorative computation of real time autocorrelation 
functions, in the classical approximation. The results for the
`plasmon' frequencies and damping rates
appear compatible with the divergencies expected from 
perturbation theory. 
}
  
\section{Motivation}
The (effectively) classical approximation has been used in 
computations of the Chern-Simons diffusion rate in the 
SU(2)-Higgs model, which is an essential ingredient in theories 
of baryogenesis. It is potentially a very useful approximation 
for non-equilibrium processes in field theory. However, its 
formal properties are not very well known yet and it is not clear 
either how well the classical field theory approximates the 
quantum theory at high temperatures. 

These issues can be studied in the weak coupling expansion. The 
classical theory has divergencies at finite temperature and the 
question naturally arises if the theory is renormalizable in some 
sense. The answer is affirmative for $\ph^4$ theory and the hot 
classical theory can approximate the quantum theory well at weak 
coupling\cite{AaSm96}. For gauge theories the 
situation is more complicated. Their leading hard thermal loop 
effects are non-local and counterterms would have to have 
this property as well (although
the specific form of the nonlocalities may 
be representable by locally coupled new degrees of 
freedom).

Still, one can suppose that any divergent quantity will be 
unobservable at low energies. Divergent frequencies and damping 
rates will be unobservable at sufficiently large time scales. 
The question is then, if there remain 
physically interesting phenomena at such time scales. 

For time independent quantities the classical approximation is a 
form of the dimensional reduction approximation. 
However, for the time component $A_0$ of 
the gauge field the coefficients of $\Tr A_0^2$ and $\Tr A_0^4$ 
are zero in the classical theory\cite{TaSm96}, while they have well defined 
values in the dimensionally reduced theory, 
including a Debije mass counterterm 
$\propto \Tr A_0^2$. The absense of this counterterm in the 
classical theory implies that the corresponding Debije mass is 
divergent. For static quantities this is not a problem since 
$A_0$ simply decouples. Such a decoupling approximation is often 
made in addition to dimensional resuction
because the perturbative Debije mass 
is relatively large. For dynamical properties one may 
hope that a similar decoupling is possible in the sense describe 
above.

The situation is more serious in the dynamical case 
since also the classical plasmon frequency associated with the 
modes $A_k$, $k=1,2,3$, is divergent\cite{Ar97}. This can be seen 
in a one loop calculation on the lattice. 
The two phenomena (diverging Debije mass 
and plasmon frequency) are related by the fact that there is no 
local counterterm available for $A_{\mu}$ in the classical 
lagrangian.

So it seems a good idea to compute nonperturbatively various 
correlation functions in the classical model and study their 
lattice spacing dependence. The results can then also be compared 
with analytic calculations, in the classical and quantum theory.
This work was undertaken with Wai Hung Tang\cite{TaSm97}.
We concentrated on autocorrelation functions
\be
C_O(t) = \langle O(t) O(0) \rangle
\ee
of gauge invariant observables with the quantum numbers of the Higgs
and $W$ fields, at zero momentum,
\be
O \to H = \vr^{\dagger}\vr,\;\;\; 
\mbox{} \to
W_{k}^{\al}= i \Tr \phd (\partial_k - iA_{k}^{\bt}\, t_{\bt}) \ph\, t_{\al},
\ee
with $\vr$ the Higgs doublet, $\ph$ its matrix version $(i\ta_2\vr^*,\vr)$
and $t_{\al} = \tau_{\al}/2$ 
the SU(2) generators. In the Higgs phase
these fields correspond to the usual Higgs and gauge fields.
Since $\ph=\sqrt{H}\, V$ with $V\in SU(2)$, $W_{\mu}^{\al}$ is
essentially the gauge field in the unitary gauge $V=1$, as
can be seen in perturbation theory by replacing $\vr$ by its vacuum
expectation value, neglecting fluctuations.
In the plasma phase this reasoning breaks down as the theory
behaves like QCD and perturbation theory runs into infrared
problems.

\section{Simulation} 

The effective classical hamiltonian on a spatial lattice can be 
given schematically in the form\cite{TaSm96}
\bea
\frac{H_{\rm eff}}{T} &=& 
\btb \sum_{\vecx}
[\half\,z_E \bar{E}^{\al}_{m\vecx} \bar{E}^{\al}_{m\vecx}
+ z_{\pi}\bar{\pi}^{\dagger}_{\vecx}\bar{\pi}_{\vecx}
+ \sum_{m<n} (1-\half\, \Tr U_{mn\vecx})
\nonumber\\ &&\mbox{}
+ (D_m\bar{\vr}_{\vecx})^{\dagger}D_m\bar{\vr}_{\vecx}
+ \lmb(\bar{\vr}^{\dagger}_{\vecx}\bar{\vr}_{\vecx} - \vb^2)^2].
\eea
We use the 
parametrization and numerical implementation of Krasnitz\cite{Kra95}. 
Comparison with 
dimensional reduction leads to the following identification of 
parameters\cite{TaSm96}
\bea
\bar\bt &\approx& \frac{4}{g^2 aT},\;\;\;
\bar \lm \approx \frac{m_H^2}{2m_W^2},\\
4\lmb\vb^2 &=& a^2 T^2 \left[ \frac{m_H^2}{T^2}
-
\frac{9g^2}{4}\,
\left(
\frac{3+\rh}{18} - \frac{3+\rh}{3}\, \frac{2\Sg}{aT}\,
\right)\, - \frac{81g^4}{16}\,
\right.
\label{VBT}\\
&& \mbox{} \left.
\times
\frac{1}{8\pi^2}\, \left(
\frac{149+9\rh}{486}
 + \frac{27 + 6\rh -\rh^2}{27}\,
\ln\frac{aT}{2} - \frac{27\et + 6\rh\bar{\et} - \rh^2
\tilde{\et}}{27} \right)\right],
\nonumber
\eea
where $a$ is the lattice spacing, $\et$, $\tilde{\et}$ and $\Sg$ are numerical
constants, $\rh = m_H^2/m_W^2$, and $g^2$, $m^2_{H,W}$, 
are running couplings and masses in the 4D 
quantum theory in the $\msbar$ scheme with scale parameter 
$\mu = 4\pi e^{-\gm} T \approx 7T$. 
Furthermore, $z_E/z_{\pi} = 1$
to a good approximation, which leaves only one parameter $z = z_E 
= z_{\pi}$ to set the time scale relative to the lattice spacing. 
Intuitively one expects $z=1 + O(g^2,\lm)$, but recent work
suggests that $z$ may have to be taken substantially 
different from 1, depending on the lattice spacing\cite{Ar97}. 

We used $z=1$, $g^2= 4/9$, $\rh = 2\lmb = 1$, and various $\btb$,
$\vb^2$ and lattice sizes $N^3$. First we did a scan with $N=24$ and
$\vb^2$ fixed, implying varying $T/m_H$ by eq.\ (\ref{VBT}). 
The ratio $T/m_H$ can be converted to $T/T_c$ since the transition 
point $T_c/m_H$ is known reasonably well. 
Then we spent more effort in the plasma phase with $(N,\btb) = (20, 11)$ 
and (32, 18.3), while $\vb^2$ was adjusted such that $T/T_c = 1.52$
in both cases.
Hence, the lattice spacing changed by a factor of 11/18.3 = 0.60 while
the physics remained the same according to (\ref{VBT}), including the
physical volume: $LT= NaT = 9N/\btb \approx \mbox{const}$. In the Higgs
phase similar effort was spent with $(N,\btb) = (20,13)$ and (32, 21.7),
without adjusting $\vb^2$, however, which led to somewhat different
$T/T_c$: 0.85 and 0.78, respectively.

\begin{figure}
\centerline{\epsfxsize=60mm\epsfbox{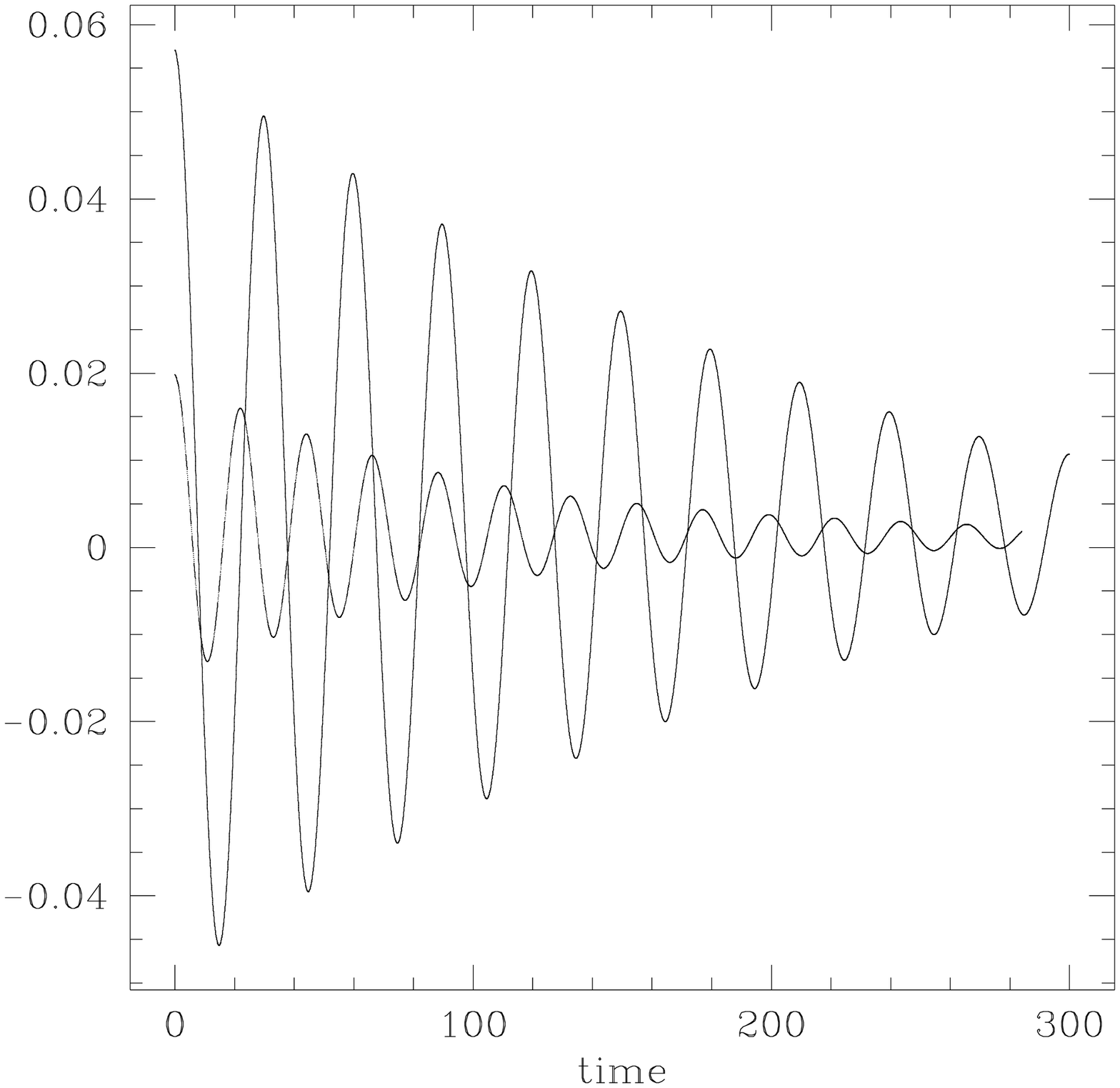}
            \epsfxsize=60mm\epsfbox{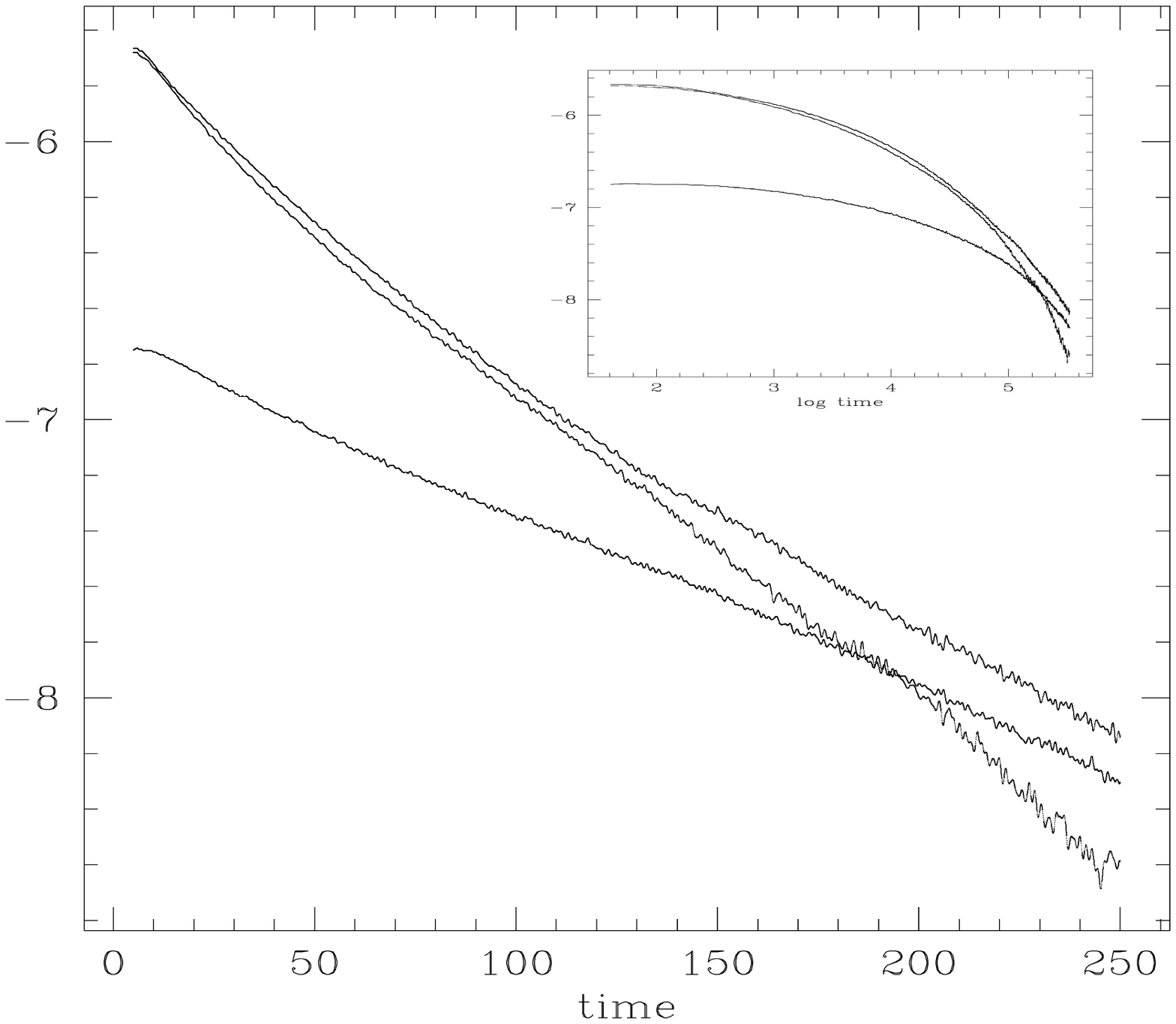}
}
\caption{Left: $C_H(t)$ in the Higgs phase for $\btb = 13$ 
(small amplitude) and 21.7 (large amplitude).
Right: plot of $\ln C_W(t)$ in the plasma phase. The crossing
lines correspond to $\btb=11$, 18.3 at $T/T_c = 1.52$.
The top curve corresponds to $\btb = 11$ at a different 
$\vb^2$ such that $T/T_c = 1.23$.
The time is in lattice units: `time' = $t/a$.
The insert shows $\ln C_w(t)$ versus $\ln t/a$.
}
\label{C}
\end{figure}

\begin{figure}
\centerline{\epsfxsize=65mm\epsfbox{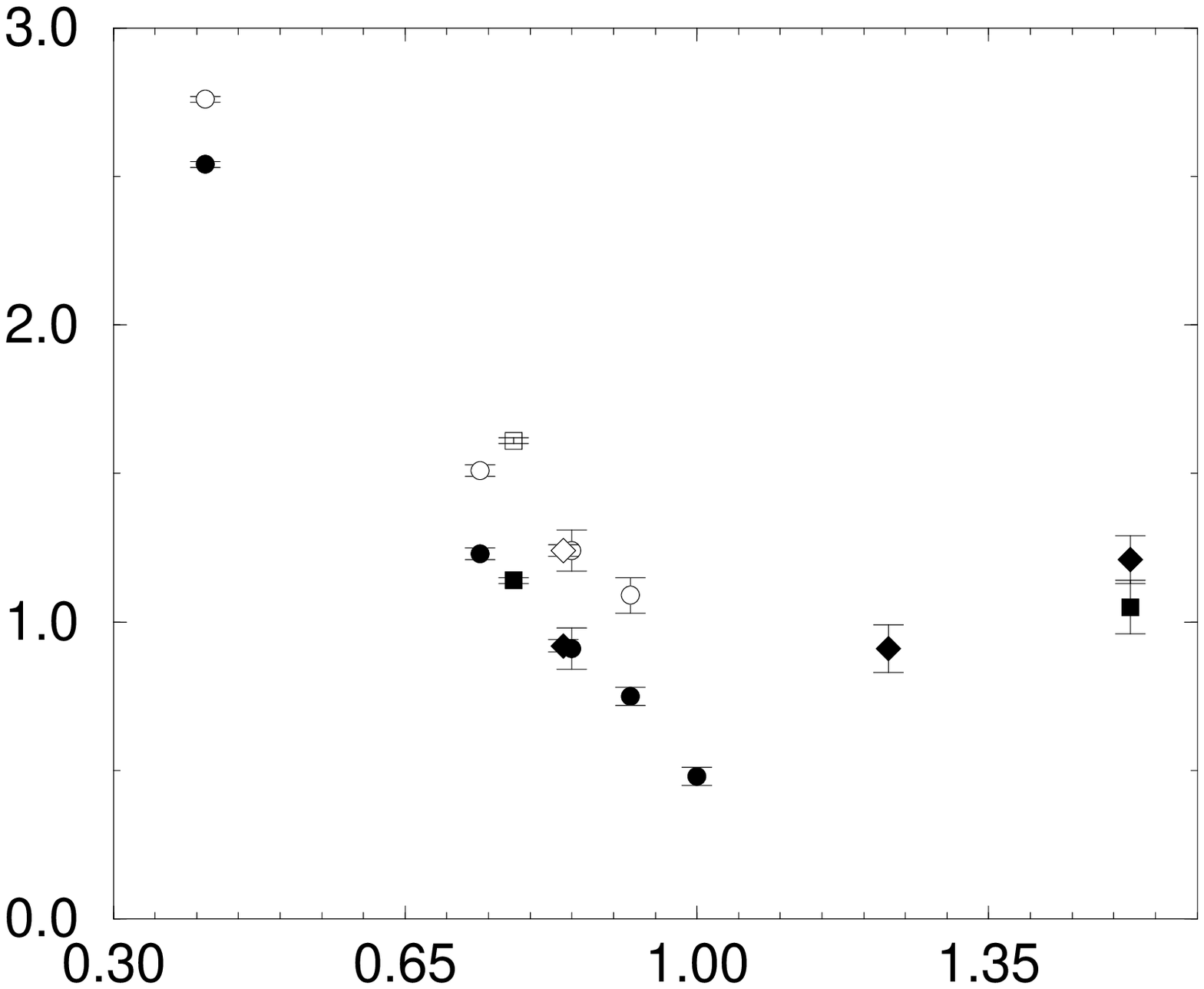}
             \epsfxsize=65mm\epsfbox{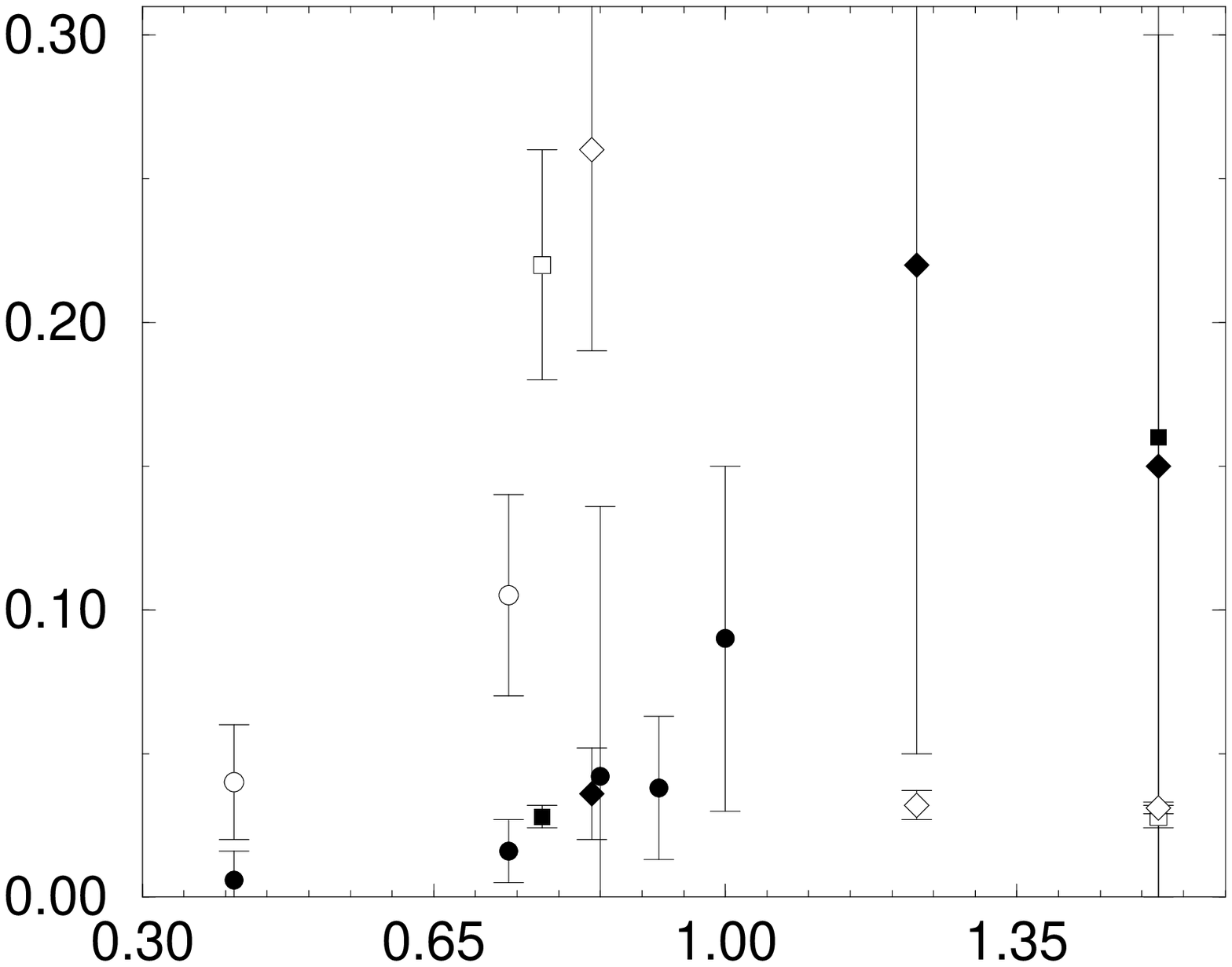}
}
\caption{
`Plasmon' frequencies $\om/g^2 T$ (left) and damping rates $\gm/g^2 T$
(right) versus $T/T_c$. 
Solid symbols correspond to $H$, open symbols to $W$; circle:
$24^3$ data; diamond: $20^3$ data; square: $32^3$ data.
}
\label{ffreqdamp}
\end{figure}

{}Fig.\ \ref{C} shows examples of correlation functions. On the left
we see $C_H(t)$ in the Higgs phase for $\btb=13$ and 21.7. The damped 
oscillations can be fitted to the pole dominance form
\be
R e^{-\gm t} \cos(\om t + \al) + \mbox{const},
\label{fit}
\ee
with $\al, \mbox{const} \approx 0$. The plot for $C_W(t)$ looks similar,
but with more damping and the additive constant effectively nonzero.
In the plasma phase the results are
qualitatively different. There $C_H(t)$ still shows oscillations with
physical ($a\om < 1$) frequencies, but with a much larger effective damping
coefficient $\gm$, and the additive background drops in time.
For $C_W(t)$ we find only a few
oscillations at unphysically short times ($t/a$ of order 2), after which
a slow roughly exponential decay sets in ($\om \to 0$ and small $\gm$
in (\ref{fit})), shown on the right in Fig.\ \ref{C}.
{}Fig.\ \ref{ffreqdamp} shows the results for $\om$ and $\gm$ 
in units of $g^2 T$ (in lattice units, all the $a\om$ and $a\gm$ are
reasonable small compared to 1).

\section{Interpretation}
Consider first the frequencies $\om_{H,W}$
in the Higgs phase, where $H$ and $W$
correspond to the usual Higgs and $W$ fields in 
leading order perturbation theory.
At first sight the small deviations of the $32^3$ data from 
the other data (interpolated by eye) suggest lattice spacing
independence. This is what we optimistically highlighted in the 
first version of the paper \cite{TaSm97}. However, our understanding
increased substantially since then and as a result (being
pressed as well by the referees of our paper)
we reanalyzed the data, in the spirit of the expected 
$1/\sqrt{a}$ divergence in $\om_W$ from one loop perturbation theory.
The upshot (expressed in the revised version)
is that the data are in fact compatible with this divergence.
A recent analysis shows that also their magnitude can
be understood from perturbation theory \cite{BoLa97}.

The damping rates on the other hand are expected to be finite to leading 
nontrivial order. In the Higgs phase they
compare well with the analytic results in the quantum
theory. For example, the $\btb=21.7$ ($T/T_c = 0.78$) data give
$\gm_W/g^2 T = 0.22(4)$ and $\gm_H/g^2 T = 0.028(4)$, to be compared with
$0.176$ and $\approx 0.02$, respectively \cite{BraPi90}.
In the plasma phase there are strong fluctations in $C_H(t)$ causing
large errors on $\gm_H$; the errors on $\gm_W$ are reasonable.  
In both phases the damping rate data are compatible with 
lattice spacing independence, which, given the large errors is not a
very strong statement. However, the $a$ independence of $\gm_W$ 
in the plasma phase ($\gm_W/g^2 T = 0.31(2), 0.028(4)$ for 
respectively $\btb = 11, 18.3$) suggests that it is a physical
quantity as well.

A more complete understanding may be expected from analysis of
the full autocorrelation functions in perturbation theory.
Work in this direction is in progress\cite{Area97}.
{}For example, the qualitative difference between $C_H(t)$ and 
$C_W(t)$ at short times can be explained this way\cite{BoLa97}.
A good understanding of the (over-?) damping in $C_W(t)$ at larger times
may also be obtainable this way. 

\section*{Acknowledgements}
We thank Gert Aarts and Alex Krasnitz  
for useful conversations. 
This work is supported by FOM and NCF/NWO.


\end{document}